\documentclass{article}
\usepackage{graphicx}  
\usepackage{amsmath}   
\usepackage[compress]{cite}
\usepackage{amssymb}   
\usepackage{bm} 
\usepackage{dcolumn}
\usepackage{color}
\usepackage{mathrsfs}
\usepackage{amsfonts}
\usepackage{varioref}
\usepackage{textcomp}

\RequirePackage[colorlinks,citecolor=blue,urlcolor=magenta,linkcolor=blue]{hyperref}
\allowdisplaybreaks
\labelformat{section}{Section #1} 
\labelformat{subsection}{Section #1} 
\labelformat{subsubsection}{Section #1}
\labelformat{subsubsubsection}{Section #1}
\labelformat{equation}{Eq.~(#1)} 
\labelformat{figure}{Fig.~#1} 
\labelformat{subfigure}{Fig.~\thefigure#1} 
\labelformat{table}{Table~#1} 
\labelformat{appendix}{Appendix #1}
\newcounter{mnotecount}[section]

\renewcommand{\themnotecount}{\thesection.\arabic{mnotecount}}

\newcommand{\mnotex}[1]
{\protect{\stepcounter{mnotecount}}$^{\mbox{\footnotesize
$
\bullet$\themnotecount}}$ \marginpar{
\raggedright\small\em
$\!\!\!\!\!\!\,\bullet$\themnotecount: #1} }

\begin{document}

\title{\bf Strong cosmic censorship conjecture for a charged AdS black hole}

\author{Chiranjeeb Singha\footnote{chiranjeeb.singha@saha.ac.in}$~^{1}$ and Naresh Dadhich\footnote{nkd@iucaa.in}$~^{2}$\\
$^{1}$\small{Theory Division, Saha Institute of Nuclear Physics, Kolkata 700064, India}\\
$^{2}$\small{IUCAA, Post Bag 4, Ganeshkhind, Pune 411007}}



\maketitle

\begin{abstract}

The strong cosmic censorship conjecture states (SCCC) that one cannot extend spacetime beyond the Cauchy horizon with a square-integrable connection. This conjecture was postulated to save the deterministic nature of the most successful theory of  gravitation, general relativity. In order to explore the validation/violation of the SCCC for the charged anti-de Sitter black hole spacetime, we compute the ratio of the imaginary part of the quasinormal mode frequencies and the surface gravity at the Cauchy horizon both analytically and numerically. The lowest value of which defines the key parameter $\beta$ determining the fate of SCCC where $\beta < 1/2$ indicates validation and else violation. We show that $\beta > 1/2$ for a charged AdS black hole with the dissipative boundary conditions in the near extremal region. Thus the SCCC is violated for this spacetime.

\end{abstract}
\section{Introduction} \label{introduction}

The loss of the deterministic nature of gravitational theories occurs for the spacetime with the Cauchy horizon (CH). This is because one cannot uniquely predict the event located at the future of CH through the evolution of an initial data. This means failure of the field equations of the gravitational theories. The presence of parameter other than mass on black hole is always accompanied by occurrence CH, for example, the Reissner-Nordstrom black hole with charge and the Kerr with rotation  \cite{Chandrasekhar:15209, Poisson:2009pwt}. This happens because gravitational attraction due to mass is opposed by repulsion due to the non-mass parameter. The CH is caused when the latter dominates over the former rendering overall gravity repulsive there.  This is at the root of the problem of the breakdown of Einstein's gravitational equation at CH. 

Fortunately, it turns out that CH is not stable against linear perturbations, and it  turns singular. The spacetime thus cannot be extended beyond it. The hope that this happens generically, in general, was termed as the strong cosmic censorship conjecture (SCCC) \cite{penrose1979singularities, Penrose:1969pc, Hawking:1970zqf}. To save the deterministic nature of the Einstein gravity, it was therefore postulated that spacetime could not be extended beyond the CH with a square-integrable connection. This SCCC version is due to Christodoulou \cite{Dafermos:2014cua, Christodoulou:2008nj}. The earlier version of the SCCC might however be violated by the Reissner-Nordstrom charged and rotating Kerr black holes \cite{Poisson1990, Ori1991, Eric2004, Bhattacharjee2016}. Whereas Christodoulou's version is always respected for the asymptotically flat black holes because the exponential blueshift at CH always dominates over the power law decay of perturbations at late time \cite{Dafermos:2014cua}. 

However, the situation changes when $\Lambda$ is included, which changes perturbations decay from power law to exponential for $\Lambda > 0$ and thereby countering the blue shift on the equal footing. That is, for asymptotically de Sitter spacetime, the late time fall-off of the perturbation modes falling into the black hole will be exponentially small, which may dominate over the exponential rise through the blueshift near CH. So, it may be possible to extend the spacetime beyond CH for asymptotically de Sitter spacetime indicating violation of SCCC \cite{PhysRevD.61.064016, Costa:2014aia, Christodoulou:2008nj}. In that case, the perturbation modes have an exponential decay, $\phi \sim \exp(-\omega_{\rm I}u)\phi_{0}$. Here $\omega_{\rm I}$ is the imaginary part of the lowest lying quasinormal modes (QNM). On the other hand, the exponential blueshift near CH will be of the form, $|\phi_{\rm cauchy}|^{2}\sim exp(\kappa_{-}u)|\phi|^{2}$ where $\kappa_{-}$ is the CH surface gravity. The determining parameter to settle the matter between the two is then the dimensionless ratio $\beta = (\omega_{\rm I}/\kappa_{-})$. That is, when $\beta < 1/2$, the perturbation modes would diverge, and the SCCC would be respected while it would be otherwise for $\beta \geq 1/2$ signaling the SCCC violation \cite{Cardoso:2017soq, Hintz:2015jkj, Dafermos:2017dbw,Hod:2020ktb,Hod:2019zoa}.

It thus seems that the SCCC may be violated for the four-dimensional asymptotically de Sitter black holes in the classical Einstein gravity, as is the case for near extremal Reissner-Nordstrom-dS (RN-dS) \cite{Cardoso:2017soq,Hod:2018dpx,Mo:2018nnu}. In contrast it is however respected for the Kerr-dS black hole \cite{Dias:2018ynt} (SCCC for a Kerr-Newman-de Sitter black hole, see \cite{Casals:2020uxa,Hod:2018lmi}). Note that the asymptotically dS spacetime admits the cosmological horizon, which creates a potential well for the infalling perturbations and that has to be climbed up and thereby losing energy resulting in decrease of $\omega_{\rm I}$. The two contrasting results for the charged and rotating black holes mean that the potential well is deeper for the Kerr-dS than that for the RN-dS; i.e., $\omega_{\rm I}$ is smaller for the former than that for the latter. That is why $\beta < 1/2$ (validation) for the rotating while it remains $> 1/2$ (violation) for the charged black hole. This is how one can understand the contrasting behavior for the asymptotically dS charged and rotating black holes  \cite{Hollands:2019whz, Bhattacharjee:2020gbo, Gwak:2018rba, Gim:2019rkl}. 

In the case of asymptotically AdS spacetime, the perturbation decay turns logarithmic from otherwise exponential if there exist stable photon orbits (SPOs) that could trap the perturbations and thereby slow down the decay process. This is what happens for the Kerr-AdS black hole, as it admits SPOs. Since exponential growth of the blueshift would always dominate over the logarithmic decay and hence the SCCC would be respected \cite{Holzegel:2011uu, Holzegel:2013kna,Kehle:2020zfg}. In the case of RN-AdS spacetime, there exists no SPO for $r_{ph} \geq r_{+}$, which would be relevant for purely the dissipative boundary condition. On the other hand, for the Dirichlet boundary condition, energy cannot dissipate through the AdS boundary, and the scalar field has to vanish at infinity. There does occur an SPO between the two horizons (CH and EH) \cite{Kehle:2018zws} which can slow down the decay, and the SCCC could then be obeyed. For the dissipative boundary condition, the energy gets absorbed by the AdS boundary \cite{Holzegel:2015swa}, then the only relevant orbit is the one occurring outside EH, which is unstable. Hence the SCCC is expected to be violated in this case. This is precisely what we wish to show in this investigation. 


This analysis also holds good for the BTZ black holes \cite{Banados:1992wn, Tang:2017enb, AKBAR2007217}, which are by construction AdS. It turns out that the charged BTZ obeys SCCC because it admits SPOs \cite{Singha:2022bvr} while the rotating one does not even when the first order back reaction of quantum fields is considered \cite{Dias:2019ery}. However, when the higher order perturbations back reaction is included, the SCCC is respected for a rotating BTZ black hole \cite{Emparan:2020rnp}. Thus asymptotically, AdS black hole in three and four dimensions is expected to respect the SCCC.


It should be noted that the existence of SPOs plays the determining role for AdS black holes. That is, the existence of SPOs with the relevant boundary conditions guarantees the SCCC validation while their non-existence the violation at the linear order perturbations. As expected the SCCC validation would however be restored when higher order perturbations back reaction is included. It is in consonance with the general belief that the linear order negative result is always overturned when the non-linear order quantum effects and/or back reaction are taken into account. We would verify the validity of this belief for RN-AdS black hole spacetime. 

This article is organised as follows: In \ref{analytical}, we present an analytical analysis for obtaining the parameter $\beta\equiv (\omega_{\rm I}/\kappa_{-})$ and comment on the status of the SCCC analytically. In \ref{numerical}, we present a detailed analysis numerically to find out the QNM frequencies and the parameter $\beta$ which ultimately decides the issue. This analysis helps us to explore the validity/violation of the SCCC numerically. This is followed by the semiclassical analysis which bears out, as expected, the general belief that violation at the classical level is always restored at semi-classical analysis. That is SCCC is validated for RN-AdS. We conclude with a discussion \footnote{ We have put $c=G=1$ throughout the paper.}.

\section{Analytical estimation of the Lyapunov exponent}\label{analytical}

In this section, we would like to compute the analytical estimation of the parameter $\beta$ for a charged AdS black hole, which is described by the metric \cite{Berti:2003ud, Wang:2004bv,Wang:2000dt, Wang:2000gsa,Konoplya:2002ky}, 
\begin{equation}\label{eq01}
d s^2 = -f(r)d t^2+ f(r)^{-1} d r^2+ r^2 d \phi^2+r^2 \sin^2\theta~;\qquad 
f(r)=\left(1-\frac{2 m}{r}+\frac{q^2}{r^2}+\frac{ r^2}{L^2}\right)~.
\end{equation}
Here $m$ and $q$ are mass and charge respectively of the black hole, and the cosmological constant $\Lambda = 3/L^2$, where $L$ is the AdS radius. 

At first, we determine the position of the photon circular orbits \cite{Chakraborty:2021dmu, Mishra:2019trb}. We note that the metric \ref{eq01} is independent of the coordinates $t$ and $\phi$, so we have two constants of motion,  energy $p_{t} = -E$ and angular momentum $p_{\phi} = \mathbb{L}$. Thus, we have only one non-trivial radial geodesic,
\begin{equation}
 \dot{r}^2=\left[E^2-f(r)\frac{\mathbb{L}^2}{r^2}\right]\equiv \left[E^2-V_{\rm eff}(r)\right]~.
\end{equation}
Here `dot' denotes derivative with respect to the affine parameter along the null geodesic, and $V_{\rm eff}(r)\equiv f(r)(\mathbb{L}^2/r^2)$ is the effective potential. Given the potential, we can immediately determine the location of the circular photon orbit $r_{\rm ph}$ by setting $V^{'}_{\rm eff} (r)$ to zero. Here `prime' denotes derivative with respect to the radial coordinate $r$. Computing $V_{\rm eff}' $ from the expression for $V_{\rm eff}$ and setting it to zero, we get the radii of the photon sphere,
\begin{eqnarray}\label{eq4}
r_{\rm ph\pm}= \frac{1}{2} \left(3 m \pm \sqrt{9 m^2-8 q^2}\right)~.
\end{eqnarray}
Here $r_{\rm ph+}$ is the radius of the outer photon sphere, which lies outside the EH, and $r_{\rm ph-}$ is the radius of the inner photon sphere, which lies between the two horizons (CH, $r_{-}$ and EH, $r_{+}$) given by the two positive roots of the quartic equation, $f(r)=0$. Now to check the stability of the photon sphere here we compute $V_{\rm eff}''$ at $r_{\rm ph \pm}$ \cite{Berry:2020ntz, Tang:2017enb}. The explicit expression of $V_{\rm eff}''$, evaluated at $r_{\rm ph\pm}$  for a charged AdS black hole spacetime, is given by,

\begin{eqnarray}
 V^{''}_{\rm eff}(r)\Big|_{r_{\rm ph\pm}}
 =- \frac{64 \mathbb{L}^2 \Big(9 m^2-8 q^2\pm 3m \sqrt{9m^2-8q^2}\Big)}{\Big(3m\pm \sqrt{9m^2-8q^2}\Big)^6} ~,
\end{eqnarray}
which is negative at the radius of the outer photon sphere and positive at the radius of the inner photon sphere. Thus, an unstable photon orbit exists for the charged anti-de Sitter black hole spacetime. The unstable photon sphere indicates that the SCCC may be violated.

 
We compute the surface gravity at CH and it is given by 
\begin{align}\label{surfacegravity}
\kappa_{-}=\frac{1}{2}f'(r_{-})=\frac{m}{r_{-}^2}-\frac{q^2}{r_{-}^3}+\frac{r_{-}}{L^2}~.
\end{align}
The Lyapunov exponent, on the other hand, is derived from the metric potential, $f(r)$ and its double derivative at the location of the photon sphere. It is then given by \cite{Mishra:2020jlw,Rahman:2018oso, Cardoso:2008bp, Cardoso:2017soq , Konoplya:2017wot,Kehle:2021ufl, Kehle:2018zws},
\begin{eqnarray}\label{eq5}
\lambda=\sqrt{\frac{f(r_{\rm ph})}{2}\left(\frac{2 f(r_{\rm ph})}{r^2_{\rm ph}}-f''(r_{\rm ph})\right)}~,
\end{eqnarray}
which for the metric \ref{eq01} gives 
\begin{equation} \label{eq6}
\lambda=\sqrt{\frac{f(r_{\rm ph})}{r^2_{\rm ph}}\Big[1- \frac{2 q^2}{r_{\rm ph}^2}\Big]}~.
\end{equation}
From the above equations, (\ref{eq6}) and (\ref{surfacegravity}), we compute the determining parameter, $\beta_{\rm ph}=(\lambda/2 \kappa_{-})$ \cite{Mishra:2020jlw, Rahman:2018oso,Cardoso:2008bp,Cardoso:2018nvb,Dias_2019,Cardoso:2017soq}, and it reads as 
\begin{eqnarray}\label{eq7}
\beta_{\rm ph}=\frac{\sqrt{\frac{f(r_{\rm ph})}{r^2_{\rm ph}}\Big[1- \frac{2 q^2}{r_{\rm ph}^2}\Big]}}{2 \big(\frac{m}{r_{-}^2}-\frac{q^2}{r_{-}^3}+\frac{r_{-}}{L^2}\big)}~.
\end{eqnarray} 
Here we have analytically computed $\beta_{\rm ph}$ for the photon sphere modes but this analysis is incomplete because the Lyapunov exponent gives the imaginary parts of the QNMs in the eikonal limit. It may however be possible that the imaginary parts of the QNMs associated with low angular momentum values are smaller than the one in the eikonal limit. So in the next section, we numerically find the QNM frequencies to find the lowest value to get the required $\beta$ which will then determine the validity/violation of the SCCC in this case.


\section{Numerical estimation }\label{numerical}

\begin{figure*}[h!]
\centering
\includegraphics[scale=0.4]{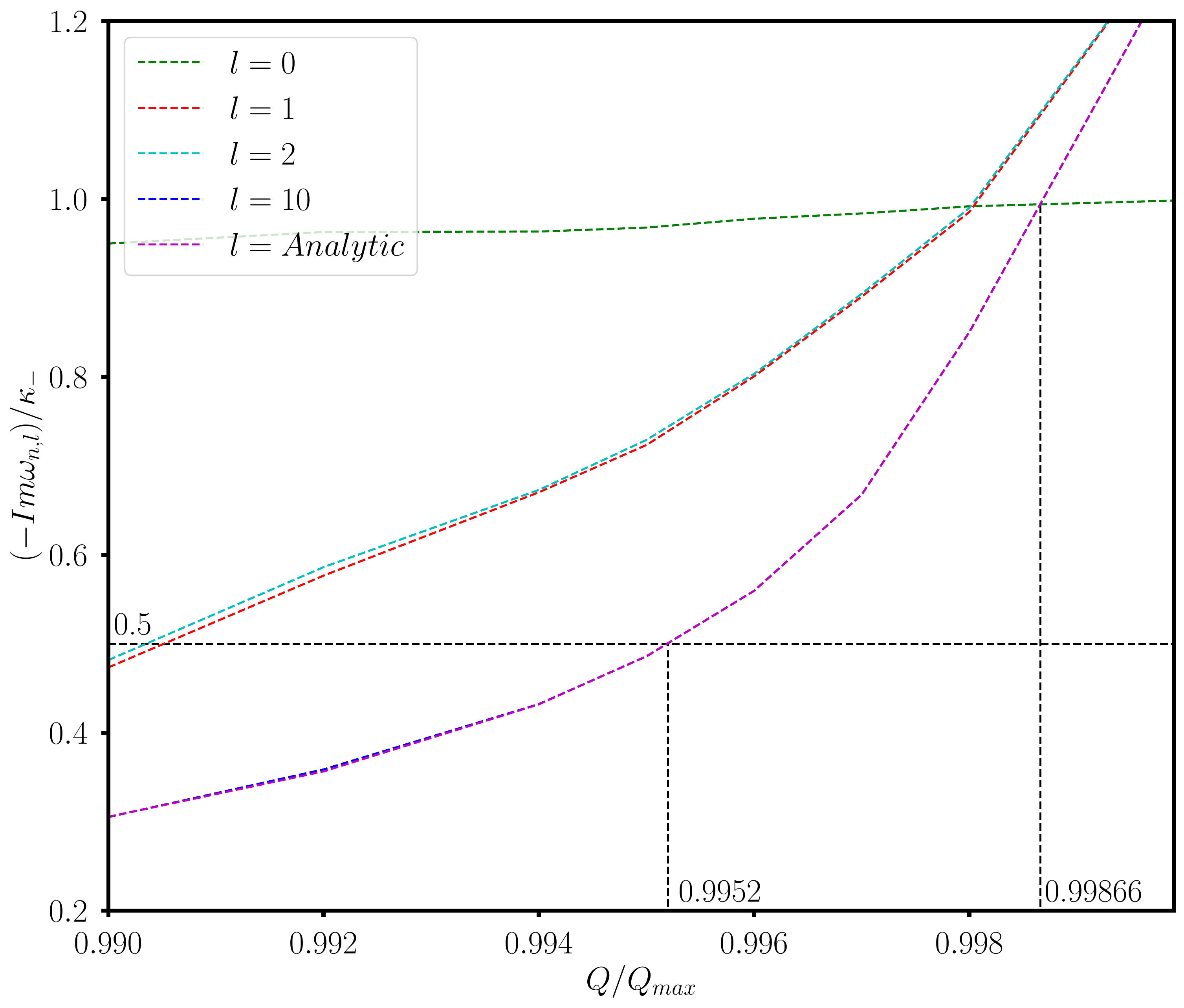}
\caption{We have plotted the ratio $\{-(\textrm{Im} \omega_{n,l})/ \kappa_{-}\}$ against $(q/q_{\rm max})$ for different choices of the angular momentum $\ell$ and for $L = 10$ and $m = 1$.  The lowest lying curve depicts $\beta$ which  is greater than the critical value (1/2) at the higher values of $(q/q_{max})$. So the violation of the SCCC occurs at near extremality. The two black dashed vertical lines indicate $(q/q_{max})$ where $\beta = 1/2$ and the $l=0$ mode becomes dominant.}
\label{figure1} 
\end{figure*}

 In this section, we will numerically find the QNM frequencies and obtain $\beta$. Here we consider the perturbation of a massless scalar field $\Phi$ on a charged AdS black hole spacetime. The evolution of the perturbation is governed by the Klein-Gordon equation $\Box \Phi=0$. We make the following ansatz for the field $\Phi$ \cite{Mishra:2020jlw} for the perturbing scalar field, as the metric is periodic in time and angular coordinates, we write 
\begin{equation}\label{eq16}
\Phi(t,r,\Omega)=\sum_{l,m} e^{-i \omega t}~\frac{\mathcal{R}(r)}{r}~Y_{lm}(\Omega)~.
\end{equation}
This ansatz gives the following master equation for the radial perturbation $\mathcal{R}(r)$,
\begin{equation}\label{eq17}
\left(\frac{\partial ^2}{\partial r_{\star}^2}+\omega^2-v_{\rm eff}(r)\right)\mathcal{R}(r)=0~,
\end{equation}
where, 
\begin{equation}
v_{\rm eff}(r)=f(r)\left(\frac{\ell(\ell+1)}{r^2}+\frac{f'(r)}{r}\right)~,
\end{equation}
is the effective potential. Here $f(r)$ is the metric function  and $dr_{\star}=\{dr/f(r)\}$ is the tortoise coordinate. We note that the effective potential, $v_{\rm eff}(r)$, diverges at infinity. The QNM frequency $\omega_{n}$ is defined as the eigenvalue of \ref{eq17}. Here we choose the following purely dissipative boundary conditions, ingoing modes at the event horizon, $r_{+}$, and, as effective potential diverges at infinity,  the mode tends to zero near the infinity, 
\begin{eqnarray}\label{eq18}
\mathcal{R}(r\to r_{+})\sim e^{-i \omega r_{\star}}~\textrm{ and}~\mathcal{R}(r\to \infty)\sim 0~.
\end{eqnarray}
Now for computing QNMs numerically, we follow the procedure and use the MATHEMATICA package developed in \cite{Cardoso:2001bb, Berti:2009kk, Cardoso:2001hn,Konoplya:2011qq}. By defining the appropriate tortoise radial coordinate and imposing the boundary conditions \ref{eq18}, we obtain the complex QNM frequencies. With the surface gravity, $\kappa_{-}$, computed analytically at CH given Eq. (5), we compute the ratio $\{-(\textrm{Im}~\omega_{n,l})/ \kappa_{-}\}$, whose minimum value would give the estimated value of the parameter $\beta$. We look for the lowest-lying QNMs in \ref{label1} for different choices of angular momentum $\ell$ and the ratio $(q/q_{\rm max})$, and plot the former against the latter for different choices of $\ell$ in \ref{figure1}. Here $q_{\rm max}$ is the extremal value, $q = m$ of the electric charge. We show that for the lowest-lying QNM, i.e., $\beta$ is $> 1/2$ in the near extremal region. Thus, the SCCC is violated for a charged AdS black hole in the near extremal region.

\begin{table*}[h!]       
\centering
\begin{tabular}{cccccccccccccccccccccccccccccccccc}         
\hline\hline            
$q/q_{max}$ && $\ell=0$ && $\ell=1$ && $\ell=2$ && $\ell=10$  && $\ell=10$ (Analytical)\\ \hline
 0.99 && 0.949904 && 0.473335 && 0.481597 && 0.305103 && 0.305374\\
 0.992 && 0.962889 && 0.576593 && 0.586118 && 0.358686 && 0.356443\\
 0.994 && 0.963403 && 0.67026 && 0.672916 && 0.432101 && 0.431853\\
 0.995 && 0.967871 && 0.72348 && 0.729168 &&  0.48586 && 0.485899 \\
 0.996 && 0.977858 && 0.800609 && 0.803455 && 0.559572 && 0.559444 \\
 0.997 && 0.983753 && 0.890227 && 0.893462 && 0.667686 && 0.667738 \\
 0.998 && 0.991792 && 0.985504 && 0.989836 && 0.850435 && 0.850368\\  
 0.999  && 0.998256 && 1.29951 && 1.29998 && 1.26532  && 1.26508\\
 
\hline\hline                                                    
\end{tabular}                                
\caption{We have presented the numerical values of the ratio $\{-(\textrm{Im} \omega_{n,l})/ \kappa_{-}\}$ for the lowest lying QNMs for different choices of the angular momentum $\ell$ and the ratio $(q/q_{\rm max})$ for the AdS radius $L = 10$ and $M = 1$. }
\label{label1}             
\end{table*}

\section{Semiclassical analysis}
In this section, we briefly discuss the higher order semiclassical analysis for charged AdS black hole. It has been recently shown that the component $T_{VV}$ of renormalized stress-energy tensor in a given state $\psi$, diverges for a wide class of theories in the Kruskal-like coordinates near CH \cite{Hollands:2019whz} as  
\begin{equation}\label{eq13}
 <T_{VV}>_{\psi}\approx \frac{C}{V^2}+t_{VV}~,   
\end{equation}
where $C$ is a constant that depends on the black hole parameters, in particular $C \sim \sqrt{m^2-q^2}$ in the case under study. Here $V$ is the Kruskal null coordinate and divergence of $t_{VV}$ would depend on $\psi$.

At first, we compute the behaviour of $t_{VV}$. For that, we consider a set of double null coordinates {U, V}, that are regular at CH. Then the quantum backreacted metric would read  \cite{Bhattacharjee2016} as 
\begin{equation}
    ds^2= \omega^2(U,V) dU dV+ r^2(U,V) d \Omega^2~.
\end{equation}
From the above metric, we get the following VV component of the semiclassical Einstein equation near CH
\begin{equation}\label{eq15}
    t_{VV}\simeq \frac{2 \partial_{V}\omega~ \partial_{V}r+\omega~\partial^2_{V}r }{r~ \omega} ~.
\end{equation}
The radial part $r (U,V)$ near CH is given \cite{Bhattacharjee2016} by 
\begin{equation}\label{radialpart}
r\approx r_{-}-\epsilon |U|+\gamma |V|^{2\beta}~, \end{equation}
where $\epsilon$ and $\gamma$ are constants of which the former is dimensionless while the latter has dimension such that the second term on the right is of length dimension. Now plugging the above expression \ref{radialpart} in the \Ref{eq15} we get
\begin{equation}
t_{VV}\approx \frac{|V|^{2 \beta-2}}{r_{-}}   
\end{equation}
which clearly diverges only when $\beta < 1/2$. For BTZ black hole $C=0$ \cite {Dias:2019ery}, and hence the divergence of $<T_{VV}>_{\psi}$ depends on the behaviour of $t_{VV}$. In dimension, $D > 3$, $C$ is non-zero \cite{Hollands:2019whz} and the first term in Eq. (13) would always be dominant as $V\to0$ at CH, and hence the renormalized energy-momentum tensor would diverge. Thus in all dimensions greater than three, SCCC would always be restored when semiclassical corrections are taken into account even if it is violated at the classical level. In particular, for the case under investigation, SCCC would be respected for the charged AdS black hole. 
\section{Discussion }\label{conclusion}

The SCCC states that the black hole spacetime cannot be extended beyond CH with the square-integrable connection \cite{penrose1979singularities, Penrose:1969pc, Hawking:1970zqf}. The validation of the SCCC saves the deterministic nature of general relativity. This conjecture is respected for all asymptotically flat black hole spacetimes. This is because the exponential growth of blueshift or mass inflation at near CH always dominates over the power law fall-off of the perturbations at the late time. 

However, this conjecture may not always be respected for asymptotically dS black holes. In that case, both the blueshift as well as the perturbations decay are exponential and hence are on an equal footing. At the late time, perturbation modes falling into the black hole will be exponentially small, which may dominate over the exponential rise in the blueshift near CH. This is indeed the case for the RN-dS black hole with $\beta > 1/2$. It is the dimensionless ratio of the imaginary part of the lowest QNM and the CH surface gravity, $\beta$, that determines the ultimate fate of SCCC, it is respected when $\beta < 1/2$ else, violated for $\beta \geq 1/2$. The former is the case for the Kerr-dS black hole for which SCCC is obeyed. What really happens is that dS spacetime has the cosmological horizon, which produces a potential well that the infalling modes have to climb out. How exponentially small the infalling modes be at late times depends upon how deep the well is. The parameter $\beta$ measures the relative dominance of the blueshift and the mode decay. It turns out that when $\beta \geq 1/2$, the well is deep enough to offset the blueshift dominance. That is indeed the case for RN-dS, for which SCCC is violated, while the opposite is true for Kerr-dS black hole. That is, in the latter case, the well is not deep enough to make the mode decay sufficiently small to offset the blueshift divergence. The opposite result in the two cases is therefore determined by how deep the potential well is at the cosmological horizon.

In the case of asymptotically AdS black holes, whenever there exist SPOs (stable photon orbits), the perturbations decay turns logarithmic, which would always be subdominant to the exponential blueshift, and hence SCCC would be respected. The logarithmic decay is due to the trapping of the modes by SPOs. If there occur no SPOs, there would be no trapping of the modes, and the decay would continue to be exponential, and SCCC would then be violated. The former is the case for the Kerr-AdS  \cite{Holzegel:2011uu, Holzegel:2013kna}  while the latter is true for RN-AdS, as we have shown in this investigation. This is also the case for the charged and rotating BTZ black holes, where the former SPOs exist and SCCC is respected, while for the latter, they do not, and it is violated. 

In the above analysis of SPOs, the boundary conditions, whether Dirichlet or dissipative, also play the decisive role in the ultimate result. All what was discussed above was for purely the dissipative boundary condition where the outer photon spherical orbit was relevant while for the Dirichlet boundary condition it is the inner orbit, and that for RN-AdS is indeed stable \cite{Kehle:2018zws}. And then SCCC is respected. Thus for RN-AdS, SCCC is violated for the purely dissipative while it is respected for the Dirichlet boundary conditions.  

Note that the role analogous to the potential well at the cosmological horizon for asymptotically dS black holes is played by SPOs for AdS black holes for which there exists no cosmological horizon. Here it is the trapping of modes by SPOs which turns the mode decay to logarithmic. Thus the existence of SPOs plays a critical determining role in the validation/violation of SCCC. The fate of SCCC for an asymptotically dS/AdS black hole is thus determined by these two properties, the depth of the potential well at the cosmological horizon for dS and the existence of SPOs for AdS. 

Here we have explored the validation/ violation of the SCCC for a charged AdS black hole spacetime both analytically and numerically. For analytical computation, we have calculated the Lyapunov exponent and surface gravity at the Cauchy horizon. We have used the relation between the Lyapunov exponent and the imaginary part of QNM frequencies in the eikonal limit. From there, we have calculated the value of the parameter $\beta$ analytically. For numerical analysis, We have derived the QNM frequencies for all possible $\ell$. From there, we have picked up the lowest QNM mode and divided it by the surface gravity at the Cauchy horizon. This gives the parameter $\beta$, which is $> 1/2$  for the charged-AdS black hole spacetime in the near extremal region. Thus the SCCC is violated.

In general, it turns out that the SCCC violation occurring at the linear order perturbations in the classical theory is always overturned at the higher order in the semi-classical theory.  \cite{Hollands:2019whz, Zilberman:2019buh}. We also carry out the higher order semi-classical analysis and show that the SCCC is indeed restored for charged AdS black hole. Thus, in this case, as well we verify the general belief that the previous negative result is always overturned when the semi-classical considerations are invoked.  

Here we have considered perturbations due to massless scalar field, it would be 
interesting to verify whether the same result holds true for the conformal scalar field or gravitational scalar field. That we leave for the future.
\section*{Acknowledgments}

We thank Christoph Kehle for bringing his work to our notice and also for very constructive criticism and helpful comments on the earlier version of the manuscript. CS thanks the Saha Institute of Nuclear Physics (SINP), Kolkata for financial support. CS also thanks Dipanjan Chakraborty for many useful discussions. 

\bibliography{mastern}

\bibliographystyle{./utphys1}

\end{document}